\documentclass[preprint, amsmath, amsymb, aps, pre,nofootinbib,showkeys]{revtex4-2}

\usepackage{graphicx}
\usepackage{dcolumn}
\usepackage{bm}

\usepackage{amsmath}
\usepackage{amssymb}
\usepackage{amsthm}
\newcommand{\dd}{{\rm d}}
\newtheorem*{remark*}{\bf Remark}

\usepackage{hyperref}
\newtheorem*{assumption*}{\bf Assumption}
\newtheorem{remark}{\bf Remark}

\makeatletter
\def\frontmatter@preabstractspace{1cm}
\makeatother

\makeatletter
\pretocmd{\frontmatter@keys@format}
  {\vfill}{}{}
\makeatother

\usepackage{setspace}
\allowdisplaybreaks

\usepackage{xcolor}

\begin{document}

\title{Generalised Langevin Dynamics:
Significance and Limitations of the Projection Operator Formalism}

\author{Christoph Widder}
\email{christoph.widder@physik.uni-freiburg.de}
\author{Tanja Schilling}%
 \email{tanja.schilling@physik.uni-freiburg.de}
\affiliation{
Institute of Physics, Albert-Ludwigs-Universität Freiburg, Hermann-Herder-Straße 3, 79104 Freiburg im Breisgau, Germany
}%




\date{\today}

\begin{abstract}
\vspace{5mm}
We discuss some mathematical aspects of the Mori-Zwanzig projection operator formalism. The core of the Mori-Zwanzig formalism is the generalised Langevin equation, which is typically derived from the Dyson-Duhamel identity. We recall the derivation of the projection operator formalism for Mori's projection by means of semigroup theory, and we discuss where rigorous methods fail for the case of Zwanzig's projection.  For bounded perturbations of the time-evolution operator (e.g. for Mori's projection), the Dyson-Duhamel identity coincides with the variation of constants formula. For unbounded perturbations (e.g. for Zwanzigs's projection), the Dyson-Duhamel identity should be considered an equation for the orthogonal dynamics, for which the existence of unique solutions has yet to be established. 
Then we recall that all properties of Mori's generalised Langevin equation follow directly from the well-posedness of Volterra equations, irrespective of the projection operator formalism. Further, we discuss the use of Mori's generalised Langevin equation as a coarse-grained model. 
Finally, we illustrate that the memory term is a coupling term that is not necessarily related to memory. To this end, we introduce projections onto subspaces of 'fast' and 'slow' variables that are associated with the spectral decomposition of skew-adjoint operators. For these projections, the memory term vanishes.
\end{abstract}

\maketitle
\newpage


\section{Introduction}
Over the past 60 years, the projection operator formalism has been established as a standard method to construct evolution equations of open systems and coarse-grained variables \cite{zwanzig2001nonequilibrium,grabert2006projection,vrugt2020projection}. 

It has been used to rationalise the Langevin-equation as an evolution equation for mesoscopic observables \cite{snook2006langevin}, it is the basis of mode coupling theory of the glass transition \cite{gotze2009complex}, it is widespread in the field of quantum mechanics of open systems \cite{breuer2002theory}, and it has been used to derive dynamic density functional theory \cite{espanol2009derivation}.

Given the wide range of research fields in which the projection operator formalism is applied one might assume that its mathematical foundations were well-established. Unfortunately, this is not the case. The method is often used in an uncontrolled manner. In this article we will dispel some common misunderstandings regarding the projection operator formalism and explain what it can be used for and where it fails. 
To this aim, we need to invoke mathematical arguments which might go beyond the common training of a physicist. We have put explanations of some mathematical terms into footnotes to facilitate the reading.


\maketitle

\subsection{Brief Summary of the Projection Operator Formalism}
Before we give a more general mathematical formulation, we briefly recall how the projection operator formalism is usually introduced in physics text books \cite{zwanzig2001nonequilibrium,grabert2006projection}.
We consider a classical system with phase space coordinates $\Gamma$ and a Hamiltonian $H(\Gamma)$. We aim to derive an effective evolution equation for some function $X(\Gamma)$ (e.g.~the generalised Langevin equation for the position of one particle in the system). 

In the first step, the equation of motion for $X(\Gamma)$ is expressed in terms of the evolution of $\Gamma$ by means of the chain rule
      \begin{equation}
        \label{eq:evolutionObservable}
        \frac{{\rm d}X_t}{{\rm d}t} =\left\{X,H\right\}(\Gamma_t) =: \mathcal{L}X(\Gamma_t) \;. \end{equation}
      Here, $\Gamma_t$ is the point in phase space reached at time $t$, we have used the notation $X_t:= X(\Gamma_t)$, the curly brackets are Poisson brackets, and we have defined the Liouvillian $\mathcal{L}$.    
      Eq.~\eqref{eq:evolutionObservable} is solved by 
      \begin{equation}
        \label{eq:formalSolution}
        X_t = e^{t\mathcal{L}}X(\Gamma_0) \; .
    \end{equation}
    Next we define a projection operator $\mathcal{P}$ on the space of phase space functions. 
    Using $\mathcal{P}$ and its orthogonal complement $\mathcal{Q}:= 1-\mathcal{P}$, we split the time-derivative of eq.~\eqref{eq:formalSolution} into two parts 
      \[
	\frac{\partial}{\partial t} e^{t\mathcal{L}}X(\Gamma{_0}) = e^{t\mathcal{L}}\mathcal{P}\mathcal{L}X(\Gamma_0) + e^{t\mathcal{L}}\mathcal{Q}\mathcal{L}X(\Gamma_0) \; .
      \]
    Then, the so-called {\it Dyson-Duhamel identity} is used \cite{kawasaki73}
    \begin{equation}
    \label{eq:Dyson1}
e^{t\mathcal{L}} = e^{t \mathcal{Q}\mathcal{L}} + \int_0^t \dd s \; e^{s\mathcal{L}}\mathcal{P}\mathcal{L}e^{ (t-s)\mathcal{Q}\mathcal{L}} 
\end{equation}
to write eq.~\eqref{eq:evolutionObservable} as 
   \begin{align}
     \frac{\dd {X}_t}{\dd t} &=
e^{t\mathcal{L}}\mathcal{P}\mathcal{L}{X}(\Gamma_0)
                                 +\int\limits_0^t\dd s\,e^{s\mathcal{L}}\mathcal{P}\mathcal{L}e^{(t-s)\mathcal{Q}\mathcal{L}}\mathcal{Q}\mathcal{L}{X}(\Gamma_0) +e^{t\mathcal{Q}\mathcal{L}}\mathcal{Q}\mathcal{L}{X}(\Gamma_0) \; .\label{eq:GLE1}
    \end{align}
    The first term on the right side is called {\it drift}. The integrand in the second term is called {\it memory term}. (This name can be misleading, as we will discuss in sec.~\ref{sec:fast_slow}.) The last term is the {\it fluctuating force}. 
   
    Depending on the choice of the projection operator $\mathcal{P}$, different versions of the generalised Langevin equation (GLE) can now be constructed, e.g.~the Mori-GLE \cite{mori1965transport}, which is linear in $X$ and therefore relatively easy to handle numerically, or the Zwanzig-GLE \cite{zwanzig2001nonequilibrium,chorin2000optimal}, which is non-linear and in which the drift is a derivative of a free energy. (We will discuss the Mori-projector in detail in sec.~\ref{sec:mori} and the Zwanzig projector in sec.~\ref{sec:zwanzig}.)
     When the projection operator approach is used to construct coarse-grained models, e.g.~in the context of molecular dynamics simulations, the fluctuating force is usually replaced by a stochastic process \cite{schilling2022coarse}. We will comment on this strategy in subsec.~\ref{sec:coarse-grained}. 
    
    In some research fields it is more common to derive a generalised Master equation rather than a GLE. For instance, in the context of open quantum systems one is usually interested in an evolution equation for the density operator rather than the observables \cite{breuer2002theory}. Regarding the procedure of the projection operator formalism this makes little difference (see e.g.~ref.\cite{grabert2006projection,seegebrecht2025concept}). 

    In the next section we will introduce the projection operator formalism once more, but in the language of semigroup theory. This will allow us to discuss the conditions under which eq.~\eqref{eq:Dyson1} holds. (Eq.~\eqref{eq:Dyson1} seems to have been taken for granted throughout the physics literature, although a bit of mathematical work is required to make sense of it \cite{givon2005,widder2025}.) We will discuss, in particular, the difference between the projection operator introduced by Mori \cite{mori1965transport} and the one by Zwanzig \cite{zwanzig1961}. Then we will recall that, actually, one can derive Mori's GLE without using the projection operator formalism at all \cite{widder2025}. We will proceed with a discussion of the predictive power of models produced by means of the projection operator formalism and, finally, we will discuss time-scale separation and the adequacy of the term "memory" for the time-convolution.   

\begin{assumption*}
    Throughout the article, let $H$ be a real or complex Hilbert space with scalar product $(.,.)$, let $\mathcal{U}(t)$ be a strongly continuous semigroup\footnote{This means: The value of an observable $x$ at time $t$ is produced by applying the operator $\mathcal{U}(t)$ to $x$, where 1) if we propagate the observable first for a duration $t\geq 0$ and then for a duration $s\geq 0$, we obtain the same value as if we propagate it directly for a duration $s+t$, and 2) under these dynamics, the observable changes continuously. } with generator $\mathcal{L}$, and let $\mathcal{P}$ and $\mathcal{Q}=1-\mathcal{P}$ be orthogonal projections on $H$. 
\end{assumption*}
    
\section{Generalised Langevin equation}\label{sec:GLE}

The observables, for which we would like to derive a GLE, will be elements of the Hilbert space $H$.  We assume that the time-evolution for vectors in $H$ is given by a strongly continuous semigroup of bounded linear operators $\mathcal{U}(t)$ with generator 
\begin{align*}
    \mathcal{L}x&:= \lim_{h\searrow0} \frac{\mathcal{U}(t)x-x}{h} \, , \\
    D(\mathcal{L}) &= \{x\in H : \mathcal{L}x \in H \} \, ,
\end{align*}
where $D(\mathcal{L})$ denotes the domain of $\mathcal{L}$. 
Since $\mathcal{U}(t)$ is strongly continuous, for all $z\in D(\mathcal{L})$ the orbit map $t\mapsto \mathcal{U}(t)z$ is continuously differentiable with derivative 
\begin{align*}
    \frac{\dd}{\dd t} \mathcal{U}(t)z &=  \mathcal{U}(t) \mathcal{L}z = \mathcal{L}\mathcal{U}(t) z \, .
\end{align*}
Furthermore, the generator $\mathcal{L}$ is densely defined\footnote{"Densely defined" means that $\mathcal{L}$ is defined only on a subspace of the Hilbert space $H$, but this subspace is dense in $H$. I.e.~we can approximate any observable in $H$ arbitrarily closely by an observable in $D(\mathcal{L})$.} and closed\footnote{
A linear operator $\mathcal{T}$ is called closed, if $\lim_n \mathcal{T} x_n = \mathcal{T}\lim_n x_n$ for all sequences $x_n \in D(\mathcal{T})$ that converge w.r.t. the graph norm $\|x_n\|_\mathcal{T}:=\|x_n\|+\|\mathcal{T}x_n\|$. 
} \cite[p.~51, Theorem~1.4]{engel}. 

In analogy to the exponential function, one often writes $\mathcal{U}(t)=e^{\mathcal{L}t}$. Mind, however, that $\mathcal{U}(t)$ cannot be represented as an exponential series in most cases. Instead, $\mathcal{U}(t)$ can be reconstructed from its generator $\mathcal{L}$ using the exponential formula for strongly continuous semigroups \cite[p.~33, Theorem 8.3]{pazy}. 

In physics text books, the generalised Langevin equation (GLE) is obtained by splitting the equation of motion by means of orthogonal projections $\mathcal{P}$ and $\mathcal{Q}=1-\mathcal{P}$. 
Written in terms of the operators that we have just introduced, eq.~\eqref{eq:GLE1} would then look like
     \begin{align*}
         \frac{\dd}{\dd t} \mathcal{U}(t) z &=\mathcal{U}(t)\mathcal{P}\mathcal{L}z + \int^t_0 \mathcal{U}(s)\mathcal{PL}e^{(t-s)\mathcal{QL}}\mathcal{QL}z  \,  {\rm d} s + e^{t\mathcal{QL}}\mathcal{QL}z \,. 
     \end{align*}
 However, to write down an equation like this we first need to check whether eq.~\eqref{eq:Dyson1} actually holds. The problem here is the following: we know -- in terms of semigroup theory -- what we mean by $e^{\mathcal{L}t}$, but it is not clear, what the expression $e^{\mathcal{QL}t}$ is supposed to mean for an arbitrary projection operator. In other words, it is unclear whether the "orthogonal dynamics" actually exists in the sense that $\mathcal{QL}$ is the generator of a strongly continuous semigroup.

A completely rigorous derivation is only possible for simple special cases. To this day, general derivations should be considered heuristic.

\subsection{Bounded perturbations}\label{sec:bounded_perturbations}

We now summarize ref.~\cite{widder2025,widder2026} to show that one can make sense of the generalized Langevin equation, if one assumes that the operator $\mathcal{\mathcal{PL}}$ is bounded, i.e. 
\begin{align*}
    \|\mathcal{PL}\| &:= \sup_{x\in D(\mathcal{PL})} \|\mathcal{PL}x\| <\infty \, .
\end{align*}
 Boundedness simplifies the problem, because it implies that the orthogonal dynamics $t\mapsto e^{t\mathcal{QL}}x=\mathcal{G}(t)x$ is well-defined and continuous for all $x\in H$. This will allow us to use semigroup theory to make sense of the Dyson-Duhamel equation eq.~\eqref{eq:Dyson1} as a variation of constants formula.
 
Since $\mathcal{L}$ is densely defined and $\mathcal{PL}$ is bounded, $\mathcal{PL}$ admits a continuous extension defined for all $x\in H$ according to
\begin{align*}
    \widehat{\mathcal{PL}}x &:= \lim_{n\to\infty} \mathcal{PL}x_n \, ,
\end{align*}
for some Cauchy sequence $D(\mathcal{PL}) \ni x_n \to x \in H$.

 From the bounded perturbation theorem \cite[p.~158, Theorem~1.3]{engel} it follows that the perturbed operator $\mathcal{\mathcal{Q}L} = \mathcal{L}-\widehat{\mathcal{PL}}=\mathcal{L}-\mathcal{PL}$ generates a strongly continuous semigroup $\mathcal{G}(t)$. (The importance of the bounded perturbation theorem in the context of the projection operator formalism has been pointed out by Zhu et al. \cite{zhu18}.)
Furthermore, we obtain the \textit{variation of constants formula} \cite[p.~161, Corollary~1.7]{engel}
\begin{equation}
\label{eq:DysonBounded}
    \mathcal{G}(t)x = \mathcal{U}(t)x - \int^t_0 \mathcal{U}(t-s)\widehat{\mathcal{PL}}\mathcal{G}(s)x  \,  {\rm d} s \, ,
\end{equation}
for all $x\in H$. In the physics literature, such a decomposition is referred to as \textit{Dyson decomposition}\cite{holian85} or \textit{Dyson-Duhamel identity}. 

The variation of constants formula (\ref{eq:DysonBounded}) immediately yields the generalized Langevin equation. I.e. for $z\in D(\mathcal{L})$ and $t\geq 0$, we have \cite{widder2026}
\begin{align}
 \frac{\dd}{\dd t} \mathcal{U}(t) z &=\mathcal{U}(t)\mathcal{P}\mathcal{L}z + \int^t_0 \mathcal{U}(t-s)\widehat{\mathcal{PL}}\mathcal{G}(s)\mathcal{QL}z  \,  {\rm d} s + \mathcal{G}(t)\mathcal{QL}z \, ,\label{equ:GLE} 
\end{align}
where the last term is the fluctuating force
\begin{align*}
    \eta(t) &:= \mathcal{G}(t)\mathcal{QL}z \, .
\end{align*}

 Thus for bounded $\mathcal{P\mathcal{L}}$  it is fine to use eq.~\eqref{eq:Dyson1}. (To see this, compare eq.~\eqref{eq:DysonBounded} and eq.~\eqref{eq:Dyson1}, note that $\mathcal{G}(t)$ corresponds to the "orthogonal dynamics" $e^{t \mathcal{Q}\mathcal{L}}$ and bring the integral to the other side.)

\begin{remark}
    We note that the fluctuating forces $\eta(t)$ are elements of the subspace $\mathcal{Q}H$. Since $\mathcal{G}(t)$ is a strongly continuous semigroup, its orbit maps are the unique \textit{mild solutions} of the associated abstract Cauchy problem \cite[p.~146, Proposition~6.4]{engel}. In other words, for all $x\in H$ the orbit map $t\mapsto \mathcal{G}(t)x$ is the unique continuous function that satisfies the equation
\begin{align*}
    \mathcal{G}(t)x-x &= \mathcal{QL}\int^t_0 \mathcal{G}(s)x \, {\rm d}s \, . 
\end{align*}
For $x\in \mathcal{Q}H$, we see that $\mathcal{G}(t)x \in \mathcal{Q}H$. This implies that $\eta(t)=\mathcal{G}(t)\mathcal{QL}z \in \mathcal{Q}H$. 
\end{remark}

\begin{remark}
    The operator $\mathcal{PL}$ is bounded for all finite-rank projections \footnote{I.e.~all projections that have a finite-dimensional range (image).} with range in $D(\mathcal{L}^\dagger)$, where $\mathcal{L}^\dagger$ is the adjoint of $\mathcal{L}$. To see this, observe that the unique continuous extension of $\mathcal{PL}$ takes the form
\begin{equation}
    \widehat{\mathcal{PL}}x = \sum^N_{n=1} (x,\mathcal{L}^\dagger z_n) z_n \, , \label{eq:BoundedP}
\end{equation}
where $\{z_n\}_n\subset D(\mathcal{L}^\dagger)$ denotes some orthonormal basis for the range of $\mathcal{P}$.

Hence it is justified to use the GLE for projection operators that are constructed via eq.~\eqref{eq:BoundedP} such as the Mori-projector \cite{mori1965transport}, $\mathcal{P}_\text{M}x:=(x,z)(z,z)^{-1}z$  with $z\in D(\mathcal{L}^\dagger)$.
However, even in this simple case, one needs to be careful about the definition of the projection, because domain issues can spoil the line of argument given above. 
\end{remark}

\begin{remark}\label{remark3}
    The assumption that $\mathcal{PL}$ is bounded implies that the range of $\mathcal{P}$ is necessarily contained in the domain $D(\mathcal{L}^\dagger)$. To see this, observe that $\mathcal{PL}\subseteq \widehat{\mathcal{PL}}$\footnote{The notation $\mathcal{A}\subseteq \mathcal{B}$ means $D(\mathcal{A})\subseteq D(\mathcal{B})$ and $\mathcal{A}x=\mathcal{B}x$ for all $x\in D(\mathcal{A})$.} implies $\widehat{\mathcal{PL}}^\dagger \subseteq [\mathcal{PL}]^\dagger$ \cite[p.~252]{reed1980}. Since $\widehat{\mathcal{PL}} \in \mathcal{B}(H)$, it follows that $\widehat{\mathcal{PL}}^\dagger \in \mathcal{B}(H)$, where $\mathcal{B}(H)$ denotes the set of bounded linear operators defined everywhere on $H$. Hence, $[\mathcal{PL}]^\dagger =\widehat{\mathcal{PL}}^\dagger$. Since $\mathcal{P}$ is an orthogonal projection and $\mathcal{L}$ is densely defined \cite[p.~51, Theorem~1.4]{engel}, we have $ [\mathcal{PL}]^\dagger = \mathcal{L}^\dagger\mathcal{P}$ \cite[p.~348, Theorem~13.2]{rudin1991}. Hence, $ \mathcal{L}^\dagger\mathcal{P} \in \mathcal{B}(H)$. This proves the assertion. This implies, for instance, that $\mathcal{PL}$ is \textit{unbounded} for the projection $\mathcal{P}x:=(x,z)(z,z)^{-1}z$ with $z\notin D(\mathcal{L}^\dagger)$. 
\end{remark}

\subsection{Unitary time-evolution}
If one intends to replace the fluctuating forces by a stochastic process in order to produce a coarse-grained model, it is important to know whether the fluctuating forces are stationary. In the following, we will show that for unitary time evolution\footnote{For the statements made in the previous subsection (sec.~\ref{sec:bounded_perturbations}), the ensemble of the microstates did not have to be stationary. This subsection now deals with the special case of equilibrium systems or systems in non-equilibrium steady-states.} the fluctuating forces are stationary in the sense that \[(\eta(t+r),\eta(s+r))=(\eta(t),\eta(s))\quad \forall r.\]  To show this, we prove that the unitarity of $\mathcal{U}(t)$ implies the unitarity of the orthogonal dynamics $\mathcal{G}(t)$.

We use arguments similar to those given for the case of Mori's projection in \cite[Corollary 2.9]{widder2025} and \cite{widder2026}. 
Recall that by Stone's theorem, $\mathcal{U}(t)$ is a unitary group if and only if $\mathcal{L}$ is \textit{skew-adjoint} \cite[p.~89, Theorem~3.24]{engel}. In this case, $\mathcal{G}(t)$ is a unitary group on the subspace $\mathcal{Q}H$: Since $\mathcal{PL}$ is bounded and $\mathcal{L}=-\mathcal{L}^\dagger$, the range of $\mathcal{P}$ is contained in the domain $D(\mathcal{L})$, cf. remark \ref{remark3}. This implies that $\mathcal{Q}x=x-\mathcal{P}x \in D(\mathcal{L})$ if and only if $x\in D(\mathcal{L})$. Hence, $D(\mathcal{LQ})=D(\mathcal{L})$. The operator $\mathcal{LQ}$ is densely defined, $\mathcal{Q}$ is an orthogonal projection and $\mathcal{L}$ is skew-adjoint, hence \cite[p.~348, Theorem~13.2]{rudin1991}
\begin{align*}
    [\mathcal{QLQ}]^\dagger &= [\mathcal{LQ}]^\dagger \mathcal{Q} = -{[\mathcal{QL}]^\dagger}^\dagger \mathcal{Q} \, .
\end{align*}
Since $\mathcal{QL}$ generates a strongly continuous semigroup, $\mathcal{QL}$ is closed \cite[p.~51, Theorem~1.4]{engel}. This implies $\mathcal{QL}={[\mathcal{QL}]^\dagger}^\dagger$ \cite[p.~252, Theorem VIII.1]{reed1980}. Hence, the operator $\mathcal{QLQ}$ is skew-adjoint:
\begin{align*}
    [\mathcal{QLQ}]^\dagger &= -\mathcal{QLQ} \, .
\end{align*}
Mind that this is a consequence of the boundedness of $\mathcal{PL}$. For arbitrary orthogonal projections $\mathcal{Q}$, the operator $\mathcal{QLQ}$ is not necessarily skew-adjoint (and also does not necessarily admit a skew-adjoint extension) \cite{givon2005}. By Stone's theorem, $\mathcal{QLQ}$ generates a unitary group, denoted $\tilde{\mathcal{G}}(t)$. The orbit maps $t\mapsto \tilde{\mathcal{G}}(t)x$ are unique mild solutions of the associated abstract Cauchy problem, hence
\begin{align*}
    \tilde{\mathcal{G}}(t)x-x &= \mathcal{QLQ}\int^t_0 \tilde{\mathcal{G}}(s)x \, {\rm d}s \, . 
\end{align*}
For $x \in \mathcal{Q}H$, this implies $\tilde{\mathcal{G}}(t)x\in \mathcal{Q}H$, hence
\begin{align*}
    \tilde{\mathcal{G}}(t)x-x &= \mathcal{QL}\int^t_0 \tilde{\mathcal{G}}(s)x \, {\rm d}s \, .
\end{align*}
For all $x\in\mathcal{Q}H$, both $t\mapsto \mathcal{G}(t)x$ and $t\mapsto \tilde{\mathcal{G}}(t)x$ are mild solutions of the same abstract Cauchy problem. Due to the uniqueness of these solutions, this implies $\tilde{\mathcal{G}}(t)x = \mathcal{G}(t)x$ for all $t\geq 0$ and $x\in \mathcal{Q}H$. This proves that the restriction of $\mathcal{G}(t)$ onto the subspace $\mathcal{Q}H$ is a unitary group on the subspace $\mathcal{Q}H$, with the understanding that we extend the semigroup $\mathcal{G}(t)$ according to $\mathcal{G}(t)x:=\tilde{\mathcal{G}}(t)x$ for all $t<0$ and $x\in \mathcal{Q}H$. In particular, this implies that the fluctuating forces $\eta(t)$ are \lq stationary\rq:
\begin{align*}
    (\eta(t+r),\eta(s+r)) &= (\mathcal{G}(t+r)\mathcal{QL}z,\mathcal{G}(s+r)\mathcal{QL}z)= (\mathcal{G}(t)\mathcal{QL}z,\mathcal{G}(s)\mathcal{QL}z) = (\eta(t),\eta(s)) \, .
\end{align*}

\subsection{Unbounded perturbations}\label{sec:unbounded_perturbations}
In this section, we discuss the case in which the orthogonal dynamics cannot be obtained from a bounded perturbation of the semigroup $\mathcal{U}(t)$. If $\mathcal{P}$ is an \textit{infinite-rank} projection, the operator $\mathcal{PL}$ is typically \textit{unbounded}. In this case, we cannot rely on the bounded perturbation theorem and its variation of constants formula. The existence of a unique mild solution of the abstract Cauchy problem 
\begin{align}
    \begin{cases}
        \frac{\dd}{\dd t} g(t,x) = \mathcal{QL}\,g(t,x) \\ 
        g(0,x) = x 
    \end{cases} \label{equ:orthogonal_dynamics_equation}
\end{align}
remains an open problem. More explicitly, the task is to find unique continuous functions $t\mapsto g(t,x)$ such that
\begin{align*}
    g(t,x)-x &= \mathcal{QL} \int^t_0 g(s,x) \,{\rm d}s \, .
\end{align*}
One might also try to resort to weak solutions in the sense of J. Ball \cite{ball} or D.~Givon and coworkers \cite{givon2005}. Notable progress has been made by D.~Givon et al. \cite{givon2005} who proved the existence of weak solutions of the orthogonal dynamics (\ref{equ:orthogonal_dynamics_equation}) for the case of stationary Hamiltonian systems. Their proof follows Friedrichs' existence proof for symmetric hyperbolic differential equations \cite{friedrichs1954}. To the best of our knowledge, the uniqueness of their solutions has not yet been addressed. 

Another way to tackle the problem is to use perturbation theorems beyond the bounded perturbation theorem such as the Desch–Schappacher and Miyadera–Voigt perturbation theorems \cite[Section III.3]{engel}. Let us give a simple example of an unbounded perturbation. Let $\mathcal{A},\mathcal{B}$ be linear operators and $D(\mathcal{A})\subseteq D(\mathcal{B})$. The $\mathcal{A}$-bound $a_0$ of $\mathcal{B}$ is defined according to
\begin{align}
   a_0 &:= \inf\left\{ a\geq 0 \,\big\vert \,  \exists\,{b\geq 0} :  \|\mathcal{B}x\| \leq b\|x\| + a\|\mathcal{A}x\| \quad \forall{x\in D(\mathcal{A})} \right\} \, .
\end{align}
If $\mathcal{A}$ generates a contraction semigroup\footnote{I.e. $\|\mathcal{S}(t)x\|\le \|x\|$ for all $x\in H$ and $t\geq 0$, where $\mathcal{S}(t)$ is the semi-group generated by $\mathcal{A}$.} on $H$ and $a_0 < 1$, then $\mathcal{A}+\mathcal{B}$ with domain $D(\mathcal{A})$ generates a strongly continuous contraction semigroup \cite[p.~173, Theorem 2.7]{engel}. 
We set $\mathcal{A}:=\mathcal{L}$ and $\mathcal{B}:=-\mathcal{PL}$. Then, for all $b\geq 0$ and $x\in D(\mathcal{L})$, we have \footnote{We note that $\|\mathcal{P}\|=1$, since $\mathcal{P}$ is an orthogonal projection. }
\begin{align*}
    \|-\mathcal{PL}x\| &\leq b\|x\| + \|\mathcal{L}x\| \, .
\end{align*}
Hence $a_0 \leq 1$, but we require the strict inequality $a_0<1$ in order obtain the desired perturbation. 

We further note that the assumption $a_0<1$ cannot be weakened in general, at least in the unitary case: If $\mathcal{\mathcal{A}}$ is essentially skew-adjoint (i.e. the closure is skew-adjoint) and $\mathcal{B}$ is skew-symmetric with $\mathcal{A}$-bound $a_0< 1$, then $\mathcal{A}+\mathcal{B}$ is essentially skew-adjoint \cite[p.~288, Theorem 4.4]{Kato1995}. Hence, by Stone's theorem, the skew-adjoint extension of $\mathcal{A}+\mathcal{B}$ generates a unitary group. However, this argument fails in general for the case $a_0=1$ \cite[p.~289]{Kato1995}.

Alternatively, one might try to find unique continuous solutions $t\mapsto g(t,x)$ of the abstract Volterra equation (Dyson identity)
\begin{align}
    g(t,x) &= \mathcal{U}(t)x - \int^t_0 \mathcal{U}(t-s) \mathcal{PL}g(s,x) \,{\rm d}s \, . \label{equ:dyson_identity}
\end{align}
For $x=\mathcal{QL}z$ such a solution would immediately yield the GLE. One possible attempt to solve this equation is to perform a \textit{vector-valued} Laplace-transform \footnote{I.e. a Laplace transform of a vector-valued function. A vector-valued function takes values in a vector space, which in our case is the Hilbert space $H$.}, cf. Arendt \cite{arendt1987}, Miller \cite{miller1975}, Grimmer \cite{grimmer1985}. 

Finally, we note that the Dyson identity is an \textit{equation}. That is, the application of the Dyson identity is equivalent to the assumption that a unique solution of eq.~\eqref{equ:dyson_identity} exists. 

\section{Mori's projection}\label{sec:mori}

The derivation of the GLE from sec.~\ref{sec:bounded_perturbations} applies to the Mori projection operator \cite{mori1965transport}
\begin{align}\label{eq:DefMori}
    \mathcal{P}_\text{M}x &:=(x,z)(z,z)^{-1}z \, ,
\end{align}
provided that $0\neq z \in D(\mathcal{L})\cap D(\mathcal{L}^\dagger)$.\footnote{Compare eq.~\eqref{eq:DefMori} to eq.~\eqref{eq:BoundedP}. To obtain the Mori-projector, the observable $z$, for which we would like to obtain the GLE, is simply used as the basis function $z_1$. In this sense, the Mori-projector is the simplest realisation of eq.~\eqref{eq:BoundedP}.} Since $\mathcal{P}_\text{M}$ is a finite-rank projection with range in $D(\mathcal{L}^\dagger)$, the operator $\mathcal{P}_\text{M}\mathcal{L}$ is densely defined and \textit{bounded}, cf. remark \ref{remark3}. The continuous extension of $\mathcal{P}_\text{M}\mathcal{L}$ is given by
\begin{align*}
    \widehat{\mathcal{P}_\text{M}\mathcal{L}}x &= (x,\mathcal{L}^\dagger z)(z,z)^{-1}z
\end{align*}
for all $x\in H$. With this, the GLE (\ref{equ:GLE}) takes the form 
\begin{align}
    \frac{{\rm d}}{{\rm d}t}\mathcal{U}(t)z &=  \omega \mathcal{U}(t)z + \int^t_0 K(s)\mathcal{U}(t-s)z  \,  {\rm d} s+ \eta(t) \, , \label{equ:GLE_Mori}\\
    \omega &:= (\mathcal{L}z,z)(z,z)^{-1} \, , \notag\\
    K(s) &:= (\eta(s),\mathcal{L}^\dagger z)(z,z)^{-1} \, , \notag
\end{align}
where $K(t)$ is called \textit{memory kernel}.

\begin{remark*}
    We emphasise that this derivation greatly simplifies the semigroup approach for the Mori projection presented in previous work \cite{widder2025}, since $\mathcal{P}_\text{M}\mathcal{L}$ is already bounded.
\end{remark*}

Note that if $\mathcal{L}$ is skew-adjoint, the drift term $\omega$ is purely imaginary, since $(\mathcal{L}z,z)^\ast= (z,\mathcal{L}z)= - (\mathcal{L}z,z)$. In this case, the drift term $\omega$ vanishes when working on a real  Hilbert space $H$. In addition, the \textit{second fluctuation dissipation theorem} (2FDT) holds, i.e. $K(s)= - (\eta(s),\eta(0))(z,z)^{-1}$. The 2FDT is a consequence of choosing the Mori projection operator. Other choices of projector do not yield the 2FDT.

\begin{assumption*}
    For simplicity, we dedicate secs.~\ref{sec:structure}-\ref{sec:coarse-grained} to the case of \textit{unitary} time-evolution on \textit{real} Hilbert spaces, so that $\omega$ from eq.~\eqref{equ:GLE_Mori} vanishes.  
\end{assumption*}

\subsection{On the structure of Mori's GLE}\label{sec:structure}

In textbooks, the GLE is introduced via the projection operator formalism, see e.g.~\cite{zwanzig2001nonequilibrium,snook2006langevin,grabert2006projection}. For Mori's projection, however, this is unnecessary. As we will show in this subsection, the GLE follows directly from the continuity of the semigroup $\mathcal{U}(t)$ irrespective of the properties of the microscopic dynamics and without any recourse to a projection or variation of constants method.
We aim to derive Mori's GLE for the case of unitary time-evolution:  For some $0\neq z \in D(\mathcal{L})$, we ask whether there exist unique functions $f(t)$ and $g(t)$ such that the equations of motion take the form 
\begin{align}
    \frac{{\rm d}}{{\rm d}t}\mathcal{U}(t)z &= \int^t_0 g(s) \mathcal{U}(t-s)z  \,{\rm d}s + f(t) \, , \label{equ:Moris_GLE} \\
    g(t) &= -(f(t), f(0))(z,z)^{-1} \, . \label{equ:2fdt}
\end{align}

 For some real-valued continuous function $g(t)$ (which will be specified in the next step), we define the vector-valued function $f(t)$ according to
\begin{align}
    f(t) &:= \frac{{\rm d}}{{\rm d}t}\mathcal{U}(t)z - \int^t_0 g(s) \mathcal{U}(t-s)z  \,{\rm d}s \, . \label{equ:fluctuating_forces}
\end{align}
Next, we impose the 2FDT:
\begin{align}
    g(t) &= -(f(t), f(0))(z,z)^{-1}  \, .
\end{align}
Inserting $f(t)$ yields a Volterra equation of the second kind for $g(t)$:
\begin{align*}
    g(t) &= -(\mathcal{U}(t)\mathcal{L}z,f(0))(z,z)^{-1} + \int^t_0 g(s) (\mathcal{U}(t-s)z,f(0))(z,z)^{-1} \,{\rm d}s \, .
\end{align*}
The coefficient functions of the Volterra equation are continuous, hence, there exists a unique continuous solution for $g(t)$ \cite[p.~25, Theorem 2.1.1]{burton}. This implies that $g(t)$ is equal to the memory kernel and the function $f(t)$ is equal to the fluctuating forces, i.e. $g(t)=K(t)$ and $f(t)=\eta(t)$. 

So far, the form of the GLE including the 2FDT follows solely from the existence and uniqueness of solutions of Volterra equations. Even the orthogonality relation $f(t) \in \mathcal{Q}H$ is established by a simple argument: Clearly $f(0)=\mathcal{L}z\in \mathcal{Q}H$, since $(\mathcal{L}z,z)= \omega (z,z)=0$. The derivative of $(f(t),z)$ is given by
\begin{align*}
    \frac{{\rm d}}{{\rm d}t}(f(t),z) &\overset{(\ref{equ:fluctuating_forces})}{=} -(f(t),\mathcal{L}z)-g(t)(z,z) \overset{(\ref{equ:2fdt})}{=} 0 \, .
\end{align*}
Hence, we conclude $f(t)\in\mathcal{Q}H$, thus $(f(t),z)=0$. Furthermore, this implies that $f(t)$ is \lq stationary\rq:
\begin{align*}
    \frac{{\rm d}}{{\rm d}r}(f(t-r),f(s-r)) &=   \frac{{\rm d}}{{\rm d}r}(\mathcal{U}(r)f(t-r),\mathcal{U}(r)f(s-r)) \\
    &\overset{(\ref{equ:fluctuating_forces})}{=} (g(t-r)\mathcal{U}(r)z,\mathcal{U}(r)f(s-r)) +(\mathcal{U}(r)f(t-r),g(s-r)\mathcal{U}(r)z) \\
    &= g(t-r)(z,f(s-r)) +g(s-r)(f(t-r),z) \\
    &= 0 \, .
\end{align*}
Hence, $(f(t+r),f(s+r))=(f(t),f(s))$. 

We have therefore established \textit{all} properties of the Mori-GLE independently of the projection operator formalism, except for the explicit representation of the fluctuating forces by means of a strongly continuous semigroup, i.e. $f(t) = \mathcal{G}(t)\mathcal{QL}z$. This derivation shows that Mori's GLE including the 2FDT does not reveal any non-trivial structure of the dynamics, and holds for any system that can be described by a strongly continuous unitary group on a real Hilbert space. 

In recent work \cite{widder2025}, we have shown that the same line of argument still applies to the non-stationary case on real or complex Hilbert spaces, for which $\mathcal{L}$ is not skew-adjoint, as well as the non-autonomous case, for which $\mathcal{L}=\mathcal{L}(t)$ is explicitly time-dependent. In the latter case, the strategy presented in this section is much easier than the projection operator formalism, as the non-autonomous case involves evolution families of the form $\mathcal{U}(t,s)$ rather than semigroups $\mathcal{U}(t)$.

\subsection{Memory-kernel and autocorrelation function}\label{sec:autocorrelation}
One might hope that the GLE provides insights regarding the dynamics of complex systems, which enable us to analyse physical phenomena such as memory effects. In general, however, it remains unclear how the different terms of the GLE can be related to physical quantities of interest. An exception is the Mori-GLE that obeys a one-to-one correspondence between the memory kernel and the autocorrelation function of the observable. \footnote{An explicit relation between the memory kernel and autocorrelation function is established by means of a Laplace transform \cite[Equation 4.11]{kubo66}.}

Applying $(\cdot,z)$ on both sides of the GLE (\ref{equ:GLE_Mori}) yields an integro-differential equation for the 'autocorrelation function' $C(t):=(\mathcal{U}(t)z,z)$,
\begin{align}
    \frac{\text{d}}{\text{d}t}C(t) &= \int^t_0 K(s) C(t-s) {\rm d}s \, , \label{equ:integro_differential}
\end{align}
where we used that $\eta(t) \in\mathcal{Q}H$, hence $(\eta(t),z)=0$. We can integrate this equation to obtain a Volterra equation of the second kind for the autocorrelation function $C(t)$,
\begin{align*}
    C(t) &= C(0) + \int^t_0 \left( \int^{t-s}_0 K(r)  \,  {\rm d}r\right) C(s) \,  {\rm d}s \, .
\end{align*}
This equation has a unique continuous solution $t\mapsto C(t)$ \cite[p.~25, Theorem 2.1.1]{burton} for any continuous memory kernel $K(t)$. Vice versa, we can differentiate eq.~\eqref{equ:integro_differential} to obtain a Volterra equation of the second kind for the memory kernel $K(t)$,
\begin{align*}
     K(t)C(0) &= C''(t)- \int^t_0 K(s) C'(t-s) {\rm d}s \, ,
\end{align*}
where $'$ denotes the derivative. Under the assumptions of sec.~\ref{sec:mori}, we have $C(0)=\|z\|^2\neq 0$ and $C(t)$ is twice continuously differentiable with second derivative $C''(t)=-(\mathcal{U}(t)\mathcal{L}z,\mathcal{L}z)$. To see this, observe that $C(t)$ is continuously differentiable with derivative $C'(t)=(\mathcal{U}(t)\mathcal{L}z,z)=-(\mathcal{U}(t)z,\mathcal{L}z)$, where we used that $\mathcal{L}$ is skew-adjoint for the case of unitary time evolution. Hence, there exists a unique continuous solution for the memory kernel $t\mapsto K(t)$ \cite[p.~25, Theorem 2.1.1]{burton}.\footnote{In practice, the autocorrelation function can usually be measured experimentally or computed in a molecular dynamics simulation, while the memory kernel is not a direct physical observable. Hence in the physics literature this one-to-one correspondence between the two functions is considered useful, in particular, in the context of developing coarse-grained simulation models. However, in subsec.~\ref{sec:coarse-grained} we will show that the Mori-GLE is not well-suited for modelling.} 
We remark that the same argument is easily applied to non-unitary time evolution on real or complex Hilbert spaces.

In practice, the memory kernel has to be evaluated numerically for a given autocorrelation function, because a direct analytical evaluation by means of the 2FDT $K(t) = -(\mathcal{G}(t)\mathcal{QL}z,\mathcal{QL}z)(z,z)^{-1}$ is typically too difficult. However, then a direct analysis of the autocorrelation function $C(t)$ appears to be simpler and less prone to numerical errors compared to the analysis of the memory kernel.

\subsection{Coarse-grained simulations}\label{sec:coarse-grained}
 If the drift, the memory kernel and the fluctuating forces can be approximated for a given set of system parameters, the GLE can serve as a coarse-grained model, i.e.~it can be integrated to obtain simulations for the observable of interest. This is a popular method in the field of computer simulation of soft materials, see e.g.~\cite{klippenstein2021introducing,li2015incorporation,izvekov2013microscopic,dalton2025memory}. In this section we ask what the predictive power of such a model is.

In the following, the Hilbert space is given by the set of real random variables $H:=L^2(\Omega,\mathcal{F},P)$ equipped with the scalar product $(x,y):=\mathbb{E}[xy]$, where $\Omega$ is the phase space, $\mathcal{F}$ is a $\sigma$-algebra and $P$ is a probability measure, cf. sec.~\ref{sec:zwanzig}.

Typically, the memory kernel $K(t)$ has to be computed numerically for a given autocorrelation function $C(t)$. In addition, we need the fluctuating forces $\eta(t)$ in order to integrate the GLE. Mind that the fluctuating force is a vector-valued function, i.e.~$\eta(t)\in H$ for all $t$. The value $\eta(t)\in H$ is a real random variable, as its initial value $\Gamma$ is determined by some outcome of the sample space (phase space) $\Omega$. Hence, the evaluation of $\eta(t)$ requires full knowledge about the initial state $\Gamma \in \Omega$. This means that the GLE is not yet an effective (coarse-grained) equation of motion, since the fluctuating forces contain the full complexity of the dynamics. In most practical cases, it is unfeasible to compute the fluctuating forces analytically (similar to the memory kernel), cf. the discussions in \cite[Section II]{hijon2009} and \cite[Section 3.3]{lin2021}. Hence, one typically resorts to a Gaussian model. The hope is that the fluctuating forces oscillate rapidly, such that they can be approximated reasonably well by \textit{coloured} Gaussian noise.\footnote{This is a non-trivial assumption as it involves the transition from deterministic to stochastic dynamics, see e.g. \cite{wong1965}\cite{givon2004}. Contrary to what is sometimes stated in the physics literature \cite{post2022molecular}, it is not simply a consequence of the central limit theorem.} This is controversial, as it is a priori unclear if there is a proper time-scale separation between the observable of interest $\mathcal{U}(t)z$ and the fluctuating forces $\eta(t)$ (cf. sec.~\ref{sec:fast_slow}), and what the actual noise would look like \cite[Section 3.3]{lin2021}.

The autocorrelation function of the fluctuating forces is given by the memory kernel, i.e. $(\eta(t),\eta(s))=-K(t-s)(z,z)$, while correlations with the initial value vanish, i.e. $(\eta(t),z)=0$. For simplicity, let us assume that $z$ is a Gaussian variable and that the mean values vanish, i.e. $\mathbb{E}[\mathcal{U}(t)z]=0$. Clearly, this implies $\mathbb{E}[\eta(t)]=0$. We can draw a realization for the initial value and the fluctuating forces $(z(\Gamma),\eta_{t_1}(\Gamma),\dots, \eta_{t_N}(\Gamma))$ from a multivariate normal distribution and integrate the GLE. By construction, the obtained coarse-grained simulations $(\tilde{z}(t_0), \dots, \tilde{z}(t_N))$ reproduce the mean values as well as the autocorrelation function of the observable of interest \cite{widder22}. (A similar line of argument holds for the non-autonomous case \cite{widder22}.)

However, since the sum of multivariate Gaussian variables is a multivariate Gaussian variable, the coarse-grained simulation $(\tilde{z}(t_0), \dots, \tilde{z}(t_N))$ is again a multivariate Gaussian variable. This implies that we must inevitably obtain the same result by drawing the coarse-grained simulation $(\tilde{z}(t_0), \dots, \tilde{z}(t_N))$ directly from a multivariate Gaussian distribution with mean zero and covariance matrix $\Sigma_{mn}=C(t_m-t_n)$, except that the latter method is less prone to numerical errors, since it does not require a numerical integration.\footnote{For non-stationary systems, the numerical stability of drawing the entire coarse-grained simulation from a multivariate Gaussian might be different compared to drawing the fluctuating forces. However, for stationary Gaussian processes, the Fourier coefficients are independent Gaussian variables \cite{berkowitz1983}. In this case, one can simply draw the respective Fourier coefficients from independent Gaussian distributions.} We thus do not gain anything by first fitting the GLE to the data. 

In any case, such methods merely reproduce the statistical quantities that have been put in. This demonstrates that the Mori-GLE cannot be used to predict the dynamics of complex systems, unless one is \textit{a priori} able to obtain reasonable approximations of the memory kernel (or autocorrelation function) as well as the distribution of initial values and fluctuating forces. 

\section{Zwanzig's projection}\label{sec:zwanzig}

In sec.~\ref{sec:bounded_perturbations}, we have established that the projection operator formalism can be treated rigorously by means of a \textit{bounded perturbation} for \textit{finite-rank} projections with domain in $D(\mathcal{L}^\dagger)$. In contrast, \textit{infinite-rank} projections typically lead to \textit{unbounded perturbations}. As an example, let us consider the Zwanzig projection operator.

The Zwanzig projection is the \textit{conditional expectation} with respect to a real random variable $u$ on the probability space $(\Omega,\mathcal{F},P)$, where $\Omega$ is the phase space, $\mathcal{F}$ is a $\sigma$-algebra and $P$ is a probability measure.\footnote{The Zwanzig projection is sometimes referred to as Chorin's projection, since Chorin et al.~were the first to formulate Zwanzig-like projections \cite{zwanzig1961} by means of conditional expectations \cite{chorin2000optimal}. Physicists usually define the Zwanzig projector as \[\mathcal{P}_ZY(\Gamma) := \frac{\int \dd \Gamma'\rho(\Gamma')\,\delta(X(\Gamma') - X(\Gamma))\,Y(\Gamma')}{\int \dd \Gamma'\rho(\Gamma')\,\delta(X(\Gamma') - X(\Gamma))}= \mathbb{E}[Y|X]\] with some phase space density $\rho$ and phase space functions $X(\Gamma),Y(\Gamma)$. 
This is the conditional expectation of $Y$ given that $X$ has the value $X(\Gamma)$ in the ensemble given by $\rho$.} 
The expectation value, scalar product and norm are defined according to 
\begin{align*}
    \mathbb{E}[x] &:= \int x \, \dd P \qquad (x,y) := \mathbb{E}[xy] \qquad \|x\| := \sqrt{(x,x)} \, .
\end{align*}
The space of measurable functions $x$ with $\|x\|<\infty$ is a Hilbert space, denoted $H=L^2(\Omega,\mathcal{F},P)$. For further reading on measure theory and probability theory, we refer to the books by Cohn \cite{cohn} and Klenke \cite{klenke}.

In classical physics, the probability measure $P$ has a density $\rho$ (phase space density) with respect to the Lebesgue measure $\mu$ on the phase space $\Omega=\mathbb{R}^n$, where $\mathcal{F}$ is either the Lebesgue algebra or the Borel algebra. One sometimes writes $P=\rho \mu$ or $\dd P= \rho \dd\mu$. The scalar product is then given by the Lebesgue integral on $\mathbb{R}^n$
\begin{align*}
   (x,y) &= \int_{\mathbb{R}^n} \rho \, x y \,  \dd\mu \, .
\end{align*}
In this case, the Hilbert space $H$ is the completion of $C^\infty_c(\mathbb{R}^n)$ w.r.t. this scalar product. 

It can be shown that the set of measurable functions of $u$ in $H=L^2(\Omega,\mathcal{F},P)$ is a closed linear subspace \cite[p.~63, Definition 2.7.3]{brockwell1991}:
\begin{align*}
    \mathcal{M}_u &:= \{ f(u) :\text{Borel functions $f$ with } f(u) \in H \} \, .
\end{align*}
The conditional expectation of $x\in H$ given $u$, denoted $\mathbb{E}[x|u]$, is defined by the orthogonal projection of $x$ onto $\mathcal{M}_u$. With this, we may introduce the Zwanzig projection according to
\begin{align*}
    \mathcal{P}_\text{Z}x &:= \mathbb{E}[x|u] \, .
\end{align*}

\subsection{Example of an unbounded perturbation}\label{ssec:unbounded_perturbation}

In this section, we give an example for which the GLE cannot be obtained from a bounded perturbation of the semigroup $\mathcal{U}(t)$. In this case, the rigorous derivation of the GLE remains an open problem, cf. sec.~\ref{sec:unbounded_perturbations}.

Let us consider the one-dimensional harmonic oscillator with Hamilton function 
\begin{align*}
    h(q,p) &= \frac{1}{2}q^2+\frac{1}{2}p^2 \, .
\end{align*}
The phase space is the two-dimensional Euclidean space $\Omega=\mathbb{R}^2$. For the $\sigma$-algebra $\mathcal{F}$, we choose the Lebesgue algebra on $\mathbb{R}^2$. We introduce the probability measure $P:\mathcal{F}\to [0,1]$ according to the Lebesgue integral
\begin{align*}
    P(V) &:= \int_V \rho \, {\rm d}\mu \, ,
\end{align*}
where $\mu$ is the Lebesgue measure on $\mathbb{R}^2$ and the \textit{phase space density} $\rho$ is given by
\begin{align*}
    \rho(q,p) &:= \frac{1}{2\pi} \exp\left\{-\frac{1}{2}q^2-\frac{1}{2}p^2\right\} \, . 
\end{align*}
With these definitions, the expectation value is given by the Lebesgue integral
\begin{align*}
    \mathbb{E}[x] &= \int x \,{\rm d}P = \int x \rho \, {\rm d}\mu = \int_{\mathbb{R}^2} x(q,p)  \rho(q,p)\,{\rm d}q {\rm d}p \, .
\end{align*}
According to Hamilton's equations of motion, the \textit{vector field} $\mathbf{F}:\mathbb{R}^2\to\mathbb{R}^2$ is given by
\begin{align*}
    \mathbf{F}(q,p) &= \begin{pmatrix}
        \partial_p H \\
        -\partial_q H 
    \end{pmatrix} = \begin{pmatrix}
        p \\
        -q  
    \end{pmatrix} \, .
\end{align*}
The vector field is continuously differentiable with bounded derivative. 
Since $\rho$ is a function of the Hamiltonian, one easily verifies that the \textit{source density} (divergence of the \textit{current density}) vanishes:
\begin{align*}
    \text{div}(\rho \mathbf{F}) &= 0 \, .
\end{align*}
This implies that the \textit{Liouville operator} (Lie derivative)
\begin{align*}
    \mathbf{F}\cdot \boldsymbol{\nabla} &= p\partial_q -q\partial_p
\end{align*}
induces a one-parameter group of linear isometries $\mathcal{U}(t)$ on $H$ \cite[Theorem 4.1]{widder2025}, i.e. $\mathcal{U}(t)$ is a unitary group on the Hilbert space $H$, cf. \cite[p.~91-92, Section 3.28]{engel}, \cite{koopman31}, \cite{spohn1975}, \cite{hunziker1968}. Its generator can be identified with the closure\footnote{A linear operator $\mathcal{T}$ is called closable, if $D(\mathcal{T}) \ni x_n\to 0$ and $\mathcal{T}x_n \to y$ implies $y=0$. Let $\mathcal{T}$ be closable, $x_n\to x$ and $\mathcal{T}x_n\to y$. Then its closure is defined by $\overline{\mathcal{T}}x:=y$.} $\mathcal{L}=\overline{\mathbf{F}\cdot \boldsymbol{\nabla}}$ in $H$ (w.r.t. the domain $D(\mathbf{F}\cdot\boldsymbol{\nabla} ) :=C^1_c(\mathbb{R}^n)$). Furthermore, for all $x\in H$, we have $\mathcal{U}(t)x=x(\varphi_t)$, where $\varphi_t$ denotes the flow generated by the vector-field $\mathbf{F}$. 
 
Let the Zwanzig projection operator be defined according to
\begin{align*}
    \mathcal{P}_\text{Z}x &:= \mathbb{E}[x|q] \, .
\end{align*}
First, we show that the operator $\mathcal{P}_\text{Z}\mathcal{L}$ is an unbounded operator. To this end, consider the sequence of observables
\begin{align*}
    x_n &:= \exp\{nq-n^2\} \, .
\end{align*}
A simple calculation shows that
\begin{align*}
    \|x_n\| = \sqrt{\mathbb{E}[x_n^2]} = 1 \, .
\end{align*}
The variables $x_n$ are Borel functions of $q$, hence, the variables $x_n$ are elements of the subspace $\mathcal{M}_q$. Thus,
\begin{align*}
    \mathcal{L}\mathcal{P}_\text{Z}x_n &= \mathcal{L}x_n = n p x_n \, .
\end{align*}
Taking the norm yields
\begin{align*}
    \|\mathcal{L}\mathcal{P}_\text{Z}x_n\| &= \sqrt{\mathbb{E}[n^2 p^2 x_n^2]} = n \, .
\end{align*}
This implies that $\mathcal{L}\mathcal{P}_\text{Z}$ is unbounded:
\begin{align*}
    \sup_{x \in D(\mathcal{L\mathcal{P}_\text{Z}})} \frac{\|\mathcal{LP}_\text{Z}x\|}{\|x\|} &\geq \sup_n \|\mathcal{LP}_\text{Z}x_n\| = \infty \, .
\end{align*}

This implies that the operator $\mathcal{P}_\text{Z}\mathcal{L}$ is also unbounded: Since $\mathcal{P}_\text{Z}$ is an orthogonal projection and $\mathcal{L}$ is skew-adjoint, we have \cite[p.~348, Theorem~13.2]{rudin1991}
\begin{align*}
    [\mathcal{P}_\text{Z}\mathcal{L}]^\dagger &= -\mathcal{L}\mathcal{P}_\text{Z} \, .
\end{align*}
However, if $\mathcal{P}_\text{Z}\mathcal{L}$ is bounded, its adjoint operator is also bounded, which contradicts the unboundedness of $\mathcal{L}\mathcal{P}_\text{Z}$. Hence, the operator $\mathcal{P}_\text{Z}\mathcal{L}$ is unbounded. 

This shows that the operator $\mathcal{L}-\mathcal{P}_\text{Z}\mathcal{L}$ is not a bounded perturbation of the generator, and we cannot apply the bounded perturbation theorem to obtain the variation of constants formula. 

\section{Fast and slow variables}\label{sec:fast_slow}
The terminology used in the literature on the GLE ("fluctuating force", "memory kernel"), suggests that $\eta(t) \in \mathcal{Q} H$ is varying rapidly compared to the variable $z\in \mathcal{P}H$, and that the memory kernel represents the dependence of the dynamics on past states. In this section, we illustrate why such interpretations require careful consideration. 

The projection operator formalism is commonly motivated by stating that the projection splits the dynamics into fast and slow (or relevant and irrelevant) parts, while a precise notion of these terms is often avoided \cite{grabert2006projection,zwanzig2001nonequilibrium,snook2006langevin}. In the following, we therefore discuss two distinct notions of slow variables, starting on the level of phase space and Hilbert space, respectively. 

Let us begin with the notion of fast and slow variables on the level of phase space. As an example, let us think of a Lennard-Jones gas in equilibrium with one heavy target particle. Our Hilbert space is then given by $L^2(\mathbb{R}^n,\mathcal{F},P)$, where $\dd P=\rho \dd \mu$, $\rho$ is the equilibrium density, and $\mu$ is the Lebesgue measure. On the level of phase space, we can then call the position $\mathbf{q}$ and momentum $\mathbf{p}$ of the target particle slow variables, and we can call the remaining variables fast. We can use Zwanzig's projection to integrate out the remaining degrees of freedom, i.e. for $f\in H$, we define
\begin{align*}
    \mathcal{P}_Z f &:= \frac{\int f \rho \, \dd \mathbf{q}_2 \cdots \dd\mathbf{q}_N\dd \mathbf{p}_2 \cdots \dd\mathbf{p}_N}{\int \rho \, \dd \mathbf{q}_2 \cdots \dd\mathbf{q}_N\dd \mathbf{p}_2 \cdots \dd\mathbf{p}_N} \, .
\end{align*}
This projection is the best approximation of a random variable $f$ in terms of a function of $(\mathbf{q},\mathbf{p})$ w.r.t. the norm of our probability space $H$. In this sense, the slow part of $f$ is given by $\mathcal{P}_Zf$, and $\mathcal{P}_Zf$ is a function of the slow variables $(\mathbf{q},\mathbf{p})$. The orthogonal part $\mathcal{Q}_Zf = f- \mathcal{P}_z f$, is 'maximally' fast in the sense that the slow part of $f$ has been subtracted. However,  
the orthogonal part $\mathcal{Q}_Zf$ generally is a function of all phase space degrees of freedom. A clear splitting of $f$ into a slow and a fast part in the above sense is only possible if $f$ is the product of a function of $(\mathbf{q},\mathbf{p})$ with a function of the remaining phase space degrees of freedom. The same issue arises when one tries to introduce a notion of fast and slow variables directly on the level of the Hilbert space as illustrated in following discussion. 

Let us attempt to formalize the idea of splitting the dynamics into rapidly and slowly varying terms for the case of \textit{unitary} time-evolution. For simplicity, we assume that $H$ is a complex Hilbert space. For real Hilbert spaces, one can simply use its complexification. Our objective is to define a suitable notion of slow variables and an orthogonal projection that projects onto the set of slow variables. In order to define such a projection, it is necessary and sufficient that the set of slow variables constitutes a closed linear subspace.\footnote{Recall that any orthogonal projection projects onto a closed linear subspace. Vice versa, any closed linear subspace admits an orthogonal projection onto this subspace. Hence, in order to define an orthogonal projection onto a set $M$, it is necessary and sufficient to assume that $M$ is a closed linear subspace.}

The time-derivative for some vector $x\in D(\mathcal{L})$ is given by
\begin{align*}
   \frac{{\rm d}}{{\rm d}t}\mathcal{U}(t)x &= \mathcal{U}(t)\mathcal{L}x \, .
\end{align*}
If $x$ is a slow variable, we expect the time-derivative to be small in some sense. A natural measure for the size of the derivative is obtained using the Hilbert space norm $\|x \|:=\sqrt{(x,x)}$. Since $\mathcal{U}(t)$ is unitary, we have $\|\mathcal{U}(t)\mathcal{L}x\|=\|\mathcal{L}x\|$. If $x$ is slow, we expect that $\|\mathcal{L}x\|$ is small. However, in order to obtain a linear subspace, we require a \textit{scale-invariant} constraint for $\|\mathcal{L}x\|$. In other words, if $x$ is a slow variable, then also $\alpha x$ must be a slow variable for all $\alpha \in \mathbb{C}$. Hence, we are lead to the condition
\begin{align}
    \|\mathcal{L}x\| &\leq \omega \|x\| \label{equ:slow_variable}
\end{align}
for some fixed frequency $\omega> 0$. Unfortunately, this condition is not enough to define a linear subspace, because it is not closed under addition.
However, if we impose similar estimates on all higher derivatives, we obtain the \textit{closed linear subspace}\footnote{These subspaces are extremely useful in spectral theory. Based on these subspaces, the first elementary proof of the spectral theorem for unbounded self-adjoint operators has been established in 1935 by Lengyel and Stone \cite{lengyel1936}, see also Leinfelder \cite{leinfelder1979}.} \cite[Lemma 1]{leinfelder1979}
\begin{align*}
    F_\omega &:= \{x\in D(\mathcal{L}^\infty): \|\mathcal{L}^nx\| \leq \omega^n \|x\| \text{ for all } n\in \mathbb{N}\} \, .
\end{align*}

The closedness of the subspace $F_\omega$ allows us to define an orthogonal projection $\mathcal{P}_\omega$ onto $F_\omega$. In addition, $F_\omega$ is an $\mathcal{L}$-invariant subspace \cite[Lemma 1]{leinfelder1979}, that is $\mathcal{L}F_\omega\subseteq F_\omega$. The Hilbert space is now given by the direct sum
\begin{align*}
    H &= \mathcal{P}_\omega H \oplus \mathcal{Q}_\omega H \, ,
\end{align*}
where $\mathcal{P}_\omega H = F_\omega$ and $\mathcal{Q}_\omega = 1-\mathcal{P}_\omega$.  A vector $x$ in $F_\omega$ obeys upper bounds for all higher derivatives. Hence, one might expect that this subspace is small. However, any vector in $H$ can be approximated by a vector in $F_\omega$ if $\omega$ is sufficiently large. In other words, $\mathcal{P}_\omega$ converges strongly towards the identity \cite[Lemma 4]{leinfelder1979}, that is
\begin{align*}
    \lim_{\omega\to\infty} \mathcal{P}_{\omega} x = x \, ,
\end{align*}
for all $x\in H$. 

In order to obtain the GLE, we have to show that $\mathcal{P}_\omega \mathcal{L}$ is bounded. Let $x\in D(\mathcal{L})$. Then,
\begin{align*}
    \|\mathcal{P}_\omega\mathcal{L}x\|^2 &= (\mathcal{P}_\omega\mathcal{L}x,\mathcal{P}_\omega\mathcal{L}x) \\
    &= -(x,\mathcal{L}\mathcal{P}_\omega\mathcal{L}x) \\
    &\leq \|x\| \|\mathcal{L}\mathcal{P}_\omega\mathcal{L}x\| \\
    &\leq \omega  \|x\| \|\mathcal{P}_\omega\mathcal{L}x\| \, .
\end{align*}
This implies $\|\mathcal{P}_\omega\mathcal{L}x\| \leq \omega  \|x\| $ for all $x\in D(\mathcal{L})$. Hence, $\mathcal{P}_\omega\mathcal{L}$ is bounded. In sec.~\ref{sec:bounded_perturbations}, we have shown that the boundedness of $\mathcal{P}_\omega\mathcal{L}$ implies the GLE:
\begin{align*}
     \frac{{\rm d}}{{\rm d}t}\mathcal{U}(t)z &= \mathcal{U}(t)\mathcal{P}_\omega\mathcal{L}z + \int^t_0\mathcal{U}(t-s)  \underbrace{\widehat{\mathcal{P_\omega L}}\mathcal{G}(s)\mathcal{Q}_\omega\mathcal{L}z}_{=0}  \,  {\rm d} s+ \mathcal{G}(t)\mathcal{Q}_\omega\mathcal{L}z \, ,
\end{align*}
where $\mathcal{G}(t)$ is the strongly continuous semigroup generated by $\mathcal{Q}_\omega\mathcal{L}$. Thus we have obtained a general strategy to split the dynamics into a fast part and a slow part and to produce a GLE for the slow variables by means of the projection operator formalism. Intriguingly however, as we will see in the next lines, this splitting implies that the memory term vanishes!  We have $D(\mathcal{LQ}_\omega)=D(\mathcal{L}-\mathcal{P}_\omega\mathcal{L})=D(\mathcal{L})$, hence $\mathcal{LQ}_\omega=\mathcal{L}-\mathcal{L}\mathcal{P}_\omega$. (Mind that for linear operators, one merely has $\mathcal{A}(\mathcal{B}+\mathcal{C})\supset \mathcal{AB}+\mathcal{AC}$ in general, but $(\mathcal{B}+\mathcal{C})\mathcal{A} = \mathcal{BA}+\mathcal{CA}$ holds \cite[p.~348]{rudin1991}.) For all $x\in D(\mathcal{L})$ and $y\in \mathcal{P}_\omega H$, we have
\begin{align*}
    (\mathcal{LQ}_\omega x, y) &= - (\mathcal{Q}_\omega x, \mathcal{L}y) = 0 \, ,
\end{align*}
where we used that the subspace $\mathcal{P}_\omega H$ is $\mathcal{L}$-invariant, hence $\mathcal{L}y \in \mathcal{P}_\omega H$. This implies $\mathcal{LQ}_\omega x \in \mathcal{Q}_\omega H$ for all $x\in D(\mathcal{L})$. In other words, we have
\begin{align*}
    \mathcal{P}_\omega\mathcal{L}\mathcal{Q}_\omega x &= 0 \quad \forall{x\in D(\mathcal{L})} \, .
\end{align*}
Since $\mathcal{P}_{\omega}\mathcal{L}\mathcal{Q}_{\omega}$ is densely defined and bounded, this implies $\widehat{\mathcal{P}_\omega\mathcal{L}}\mathcal{Q}_\omega=0$. Since $\mathcal{G}(s)\mathcal{Q}_\omega\mathcal{L}z \in \mathcal{Q}_\omega H$, this shows that the memory term indeed vanishes. Furthermore, the variation of constants formula is reduced to
\begin{align*}
    \mathcal{G}(t)x &= \mathcal{U}(t)x
\end{align*}
for all $x\in \mathcal{Q}_\omega H$. This implies that $\mathcal{Q}_\omega H$ is $\mathcal{U}(t)$-invariant. Therefore, the restriction of $\mathcal{U}(t)$ onto $\mathcal{Q}_\omega H$ is again a strongly continuous semigroup on the Hilbert space $\mathcal{Q}_\omega H$ \cite[p.~43 ]{engel}. Its generator is the restriction of $\mathcal{L}$ onto the domain $D(\mathcal{L})\cap \mathcal{Q}_\omega H$  \cite[p.~61]{engel}. Such semigroups are called \textit{subspace semigroups}. Hence, the restriction of $\mathcal{G}(t)$ onto the subspace $\mathcal{Q}_\omega H$ is simply the subspace semigroup of $\mathcal{U}(t)$ for the subspace $\mathcal{Q}_\omega H$. 

From a mathematical point of view, this result is a trivial special case that occurs if the subspace $\mathcal{Q}H$ is $\mathcal{U}(t)$-invariant. In this case, it is trivial to show that the memory term vanishes, irrespective of the unitarity of the semigroup $\mathcal{U}(t)$ or the boundedness of $\mathcal{PL}$.

The proposed splitting into rapidly and slowly varying terms resembles a spectral decomposition. We remark that we required the skew-adjointness of $\mathcal{L}$ on a complex Hilbert space $H$ in order to define the closed linear subspaces $F_\omega$. In the non-unitary case, where $\mathcal{L}$ is typically not even a normal operator, we do not have the spectral theory required in order to perform such a decomposition.

From a physical point of view, it is important to consider the implications for the interpretation of the different terms of the GLE. 
If we assume unitary time evolution and if we split the dynamics into slow and fast contributions by means of a projection onto $F_\omega$, there is no 'memory term' or 'friction term' that might reveal any physical properties. In this case, the subspaces $\mathcal{P}_\omega H$ and $\mathcal{Q}_\omega H$ do not couple, and the memory term vanishes. If one deviates from this decomposition, the subspaces $\mathcal{P}H$ and $\mathcal{Q} H$ are typically coupled, and a memory term appears. This means that the memory term is a \textit{coupling term} that compensates for the fact that the subspace $\mathcal{Q}H$ is typically not $\mathcal{U}(t)$-invariant. 

One might rightfully argue that $F_\omega$ might not yield a suitable notion of a slow variable. In addition, one would like the orthogonal complement of the set of slow variables to resemble fast variables. It appears to be difficult, however, to find other definitions of slow variables that give rise to closed linear subspaces.

\section{Summary and Conclusion}
We hope to have resolved some common misconceptions regarding the projection operator formalism and the generalised Langevin equation. 

Textbook-derivations of the projection operator formalism typically utilise a variation of constants formula, which is called {\it Dyson-Duhamel identity}.  We discussed that this formula can be proven by means of the bounded perturbation theorem, if $\mathcal{PL}$ is bounded (e.g.~for the Mori-projector), while it has not yet been proven for unbounded $\mathcal{PL}$ (e.g.~the Zwanzig-projector).  As an example for which the orthogonal dynamics fails to be a bounded perturbation of the time-evolution operator, we applied the Zwanzig-projector to the harmonic oscillator. In a general context, the Dyson-Duhamel identity should be considered an \textit{equation} for the orthogonal dynamics rather than an identity, as it remains unknown how the orthogonal dynamics can be defined in first place. 

We then recalled that all properties of Mori's GLE can actually be derived without using the projection operator formalism at all. They follow directly from the well-posedness of Volterra equations and basic calculus. The fluctuation dissipation theorem then holds by construction, and should be viewed as the defining property of the memory kernel rather than a theorem. 

Further we  argued that the Mori-GLE is not of much use if one intends to construct a coarse-grained model, unless one is able to obtain a priori estimates of the different terms. One can as well sample the entire trajectory from a simple distribution, rather than first compute the memory kernel from the autocorrelation function and then propagate the GLE using Gaussian fluctuating forces.

Finally, we illustrated that the memory term should be considered a \textit{coupling term} that is not necessarily related to memory. The main idea of the projection operator formalism is to split the dynamics into slowly and rapidly varying terms by means of an orthogonal projection. However, it appears to be difficult to provide a reasonable decomposition of the Hilbert space into subspaces of slow and fast variables. The closest definition for such a decomposition that we were able to provide makes use of the subspaces associated with the spectral decomposition of skew-adjoint operators. However, these subspaces are invariant under time-evolution, hence, they do not couple. As a consequence, the memory term vanishes. 

\begin{acknowledgments}
     We acknowledge funding by the German Research Foundation (DFG) in Project 535083866. We thank Martin Oettel, Jens Weimar, Nico van der Vegt, Niklas Wolf and Fabian Koch for helpful discussions.
\end{acknowledgments}

\appendix


\bibliographystyle{unsrt-nocase}
\bibliography{refs}

\end{document}